\documentclass[aps,prb,twocolumn,amsmath,superscriptaddress]{revtex4}
\usepackage{graphicx}
\usepackage{color}
\usepackage{epstopdf}
\usepackage{xcolor}

\begin{document}

\title{Strain induced coherent dynamics of coupled carriers and Mn spins in a quantum dot}

\author{A. Lafuente-Sampietro}
\affiliation{Universit\'{e} Grenoble Alpes, Institut N\'{e}el, F-38000 Grenoble, France}
\affiliation{CNRS, Institut N\'{e}el, F-38000 Grenoble, France}

\author{H. Boukari}
\affiliation{Universit\'{e} Grenoble Alpes, Institut N\'{e}el, F-38000 Grenoble, France}
\affiliation{CNRS, Institut N\'{e}el, F-38000 Grenoble, France}

\author{L. Besombes}\email{lucien.besombes@grenoble.cnrs.fr}
\affiliation{Universit\'{e} Grenoble Alpes, Institut N\'{e}el, F-38000 Grenoble, France}
\affiliation{CNRS, Institut N\'{e}el, F-38000 Grenoble, France}

\date{\today}

\begin{abstract}

We report on the coherent dynamics of the spin of an individual magnetic atom coupled to carriers in a semiconductor quantum dot which has been investigated by resonant photoluminescence of the positively charged exciton (X$^{+}$). We demonstrate that a positively charged CdTe/ZnTe quantum dot doped with a single Mn atom forms an ensemble of optical $\Lambda$ systems which can be addressed independently. We show that the spin dynamics of the X$^{+}$-Mn complex is dominated by the electron-Mn exchange interaction and report on the coherent dynamics of the electron-Mn spin system that is directly observed in the time domain. Quantum beats reflecting the coherent transfer of population between electron-Mn spin states, which are mixed by an anisotropic strain in the plane of the quantum dot, are clearly observed. We finally highlight that this strain induced coherent coupling is tunable with an external magnetic field.

\end{abstract}

\maketitle

Semiconductor quantum dots (QDs) are solid-state systems which permit efficient manipulation of single charge and spin. The optical properties of a QD can be used to control the spin state of carriers, nuclei as well as individual \cite{Besombes2004,LeGall2011,Kudelski2007,Kobak2014} or pairs \cite{Besombes2012,Krebs2013} of magnetic atoms. The spin of one magnetic atom in a QD can be prepared by the injection of spin polarized carriers and its state can be read through the energy and polarization of the photons emitted by the QD. The insertion of a magnetic atom in a QD where the charge and strain states can be controlled offers additional degrees of freedom to tune the properties of the localized spin. Development of information processing using the spin of magnetic atoms in semiconductors will require tuning the coherent coupling between two or more spins. Controlling the exchange coupling of magnetic atoms with the spin of a carrier is an attractive approach to coherently transfer information between the atomic localized spins.

Here we report on the coherent dynamics of the spin of a magnetic atom coupled to the spin of individual carriers. This dynamics has been investigated by resonant photoluminescence (PL) of the positively charged exciton (X$^{+}$) coupled to a single Mn in a p-doped CdTe/ZnTe QD \cite{Varghese2014}. We have analysed the spin dynamics of the X$^+$-Mn complex under resonant excitation and show that it is controlled by the electron-Mn (e-Mn) interaction. We demonstrate that a positively charged Mn-doped QD forms optical $\Lambda$ systems that can be addressed with resonant lasers. We show that the coherent dynamics of the coupled electron and Mn spins is extremely sensitive to the local strain at the Mn location. Oscillations in the circular polarization rate of the resonant PL reveal quantum beats between e-Mn spin states mixed by an anisotropic strain in the plane of the QD. This coherent coupling, characterized by a change in the angular momentum $\Delta S_z=\pm2$, can be tuned by a magnetic field.

\begin{figure}[hbt]
\includegraphics[width=3.25in]{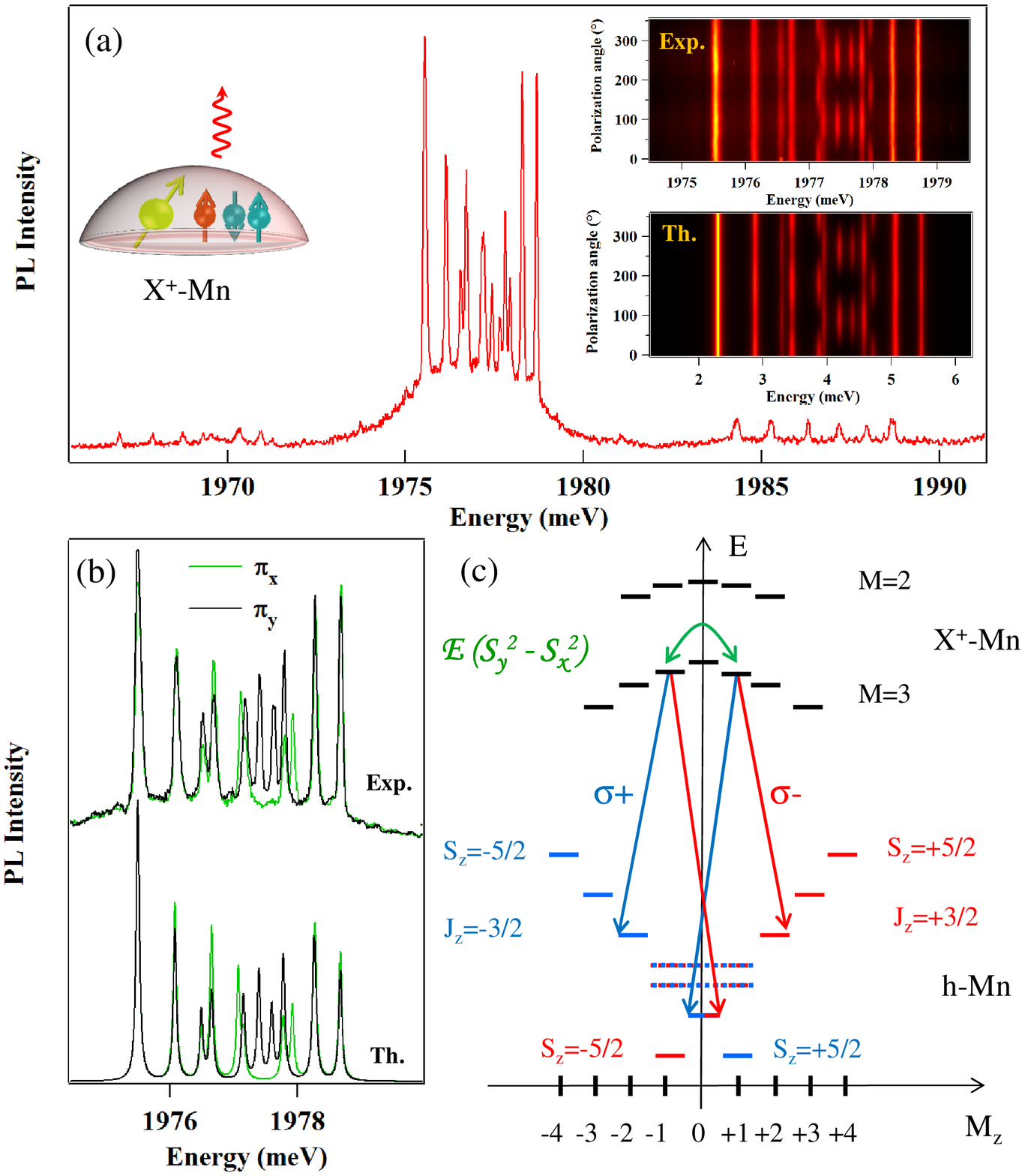}
\caption{(a) PL at T=5K of a positively charged Mn-doped CdTe QD. Inset: experimental and calculated intensity maps of the linear polarization dependence of the PL from $X^+$-Mn. (b) Experimental (Exp.) and calculated (Th.) spectra from $X^+$-Mn in linear polarizations. The parameters in the calculation are: $I_{hMn}=345 \mu eV$, $I_{eMn}=-175\mu eV$, $\rho_s/\Delta_{lh}=0.09$, $\theta$=0 and $\eta=30\mu eV$. (c) Energy levels of the ground (h-Mn) and excited ($X^+$-Mn) states as a function of their angular momentum (M$_z$). Optical $\Lambda$ systems associated with the e-Mn states $|3,+1\rangle$ and $|3,-1\rangle$, coupled by the strain anisotropy $E(S_y^2-S_x^2)$, are presented. The levels in dotted lines corresponds to the h-Mn states $|-1/2\rangle|\Uparrow\rangle$ and $|+1/2\rangle|\Downarrow\rangle$ coupled by the valence band mixing. Optical recombination towards these levels leads to the linearly polarized lines observed in (b).}
\label{Fig1}
\end{figure}

The sample consist of Mn-doped CdTe QDs grown by molecular beam epitaxy on a p-doped ZnTe(001) substrate\cite{Wojnar2011}. A positive bias voltage is applied between a 6 nm thick gold semi-transparent Schottcky gate deposited on the surface of the sample and the substrate in order to trap a hole in the QD \cite{Varghese2014}. Individual QDs containing one Mn are isolated using micro-spectroscopy techniques \cite{LeGall2010}.

When a hole is trapped in a QD with a single Mn, one has to consider the hole-Mn (h-Mn) exchange interaction described by the spin Hamiltonian ${\cal H}_{hMn}^{ex}=I_{hMn}\vec{S}\cdot\vec{J}$. Here, I$_{hMn}$ is the exchange integral of the hole with the Mn ($S=5/2$) and $\vec{J}$ is the hole spin operator. $\vec{J}$, represented in the basis of the two low energy heavy-hole states, is related to the Pauli matrices $\tau$ by $J_z= \frac{3}{2}\tau_z$ and $J_{\pm}= \xi \tau_{\pm}$ with $\xi=-2\sqrt{3}e^{-2i\theta}\rho_s/\Delta_{lh}$. $\rho_s$ is the coupling energy between heavy and light holes split by the energy $\Delta_{lh}$ and $\theta$ is the angle relative to the [110] axis of the principal axis of the anisotropy responsible for a valence band mixing \cite{Fernandez2006,Leger2007}. The h-Mn complex forms a spin ladder with a quantization axis along the growth direction and a ground state, twice degenerated, with opposite spin orientation of the hole and the Mn (M$_z$=$\pm$1 in Fig.~\ref{Fig1}(c)).

\begin{figure}[hbt]
\includegraphics[width=3.3in]{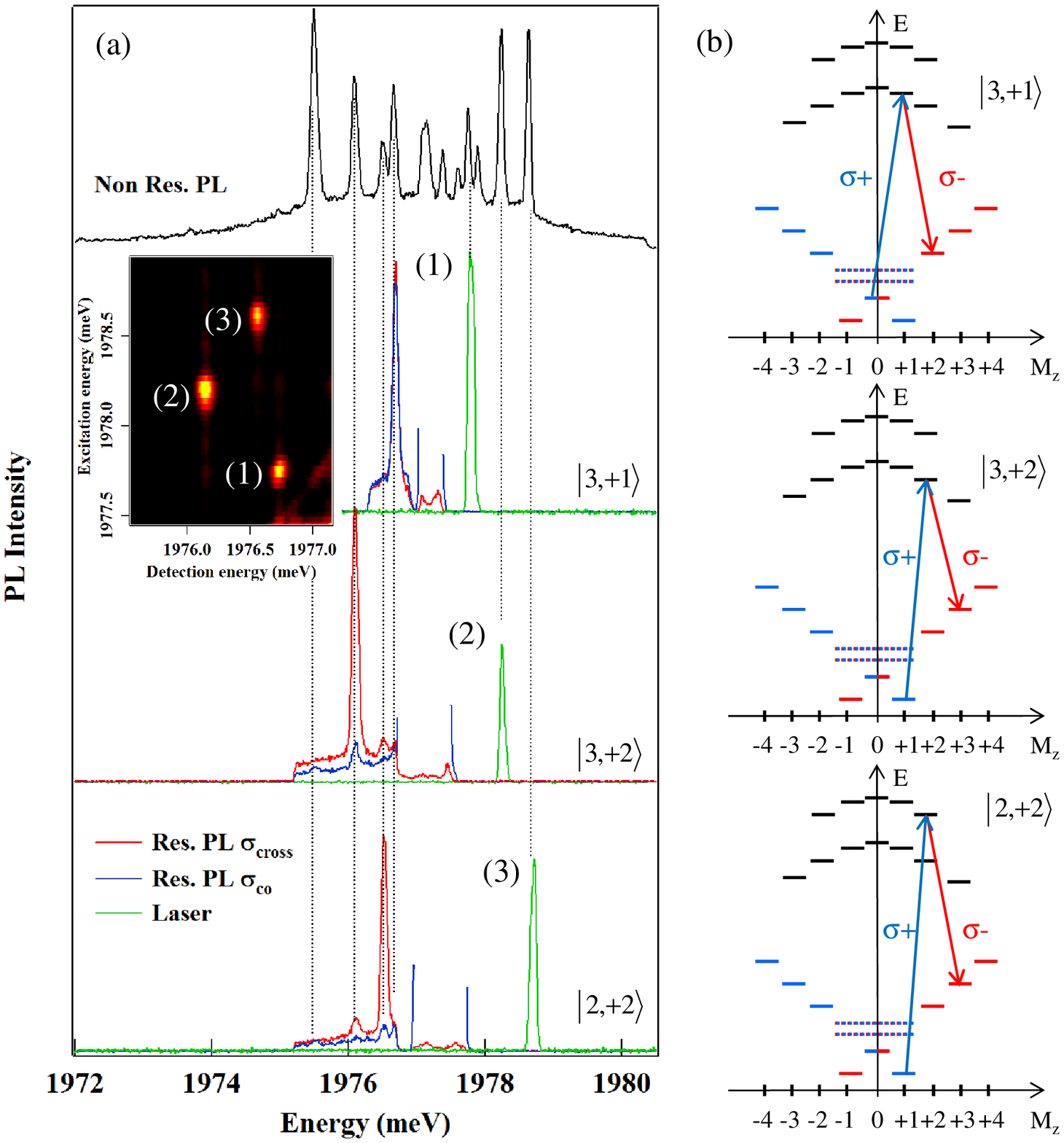}
\caption{((a) Non resonant (Non Res.) and resonant (Res.) PL of X$^+$-Mn. Co and cross circularly polarized PL spectra are collected for three different energies of the CW resonant laser (green). Inset: intensity map of the cross-circularly polarized PL detected on the low energy side of X$^+$-Mn as the CW laser is scanned through the high energy side. (b) Energy levels of X$^+$-Mn and identification of the three resonances observed in (a) corresponding to the optical $\Lambda$ systems associated with the e-Mn states $|3,+1\rangle$, $|3,+2\rangle$ and $|2,+2\rangle$.}
\label{Fig2}
\end{figure}

When an exciton is injected in the QD loaded with a hole, one has to consider the X$^+$-Mn complex. For X$^+$ in its ground singlet state, both holes have opposite spins and the structure of X$^+$-Mn is dominated by the electron-Mn exchange interaction described by the Hamiltonian ${\cal H}_{eMn}^{ex}=I_{eMn}\vec{S}\cdot\vec{\sigma}$, with $\vec{\sigma}$ the electron spin and $I_{eMn}$ the exchange integral of the electron with the Mn. The twelve e-Mn states are split into a ground state sextuplet (total spin M=3) and a fivefold degenerated manifold (total spin M=2) (Fig.~\ref{Fig1}(c)). These energy levels are labelled $|M,M_z\rangle$.

To describe the details of the PL of the X$^+$-Mn and its spin dynamics, we have taken into account the perturbation of the wave function of the hole by its exchange coupling with the Mn \cite{Besombes2005,Trojnar2013,Besombes2014}. This perturbation depends on the hole-Mn exchange energy and can be represented by an effective spin Hamiltonian ${\cal H}_{scat}=-\eta S_z^2$ with $\eta>$0. It is twice for X$^+$-Mn where two holes interact with the Mn and consequently affects the energy of the optical recombination to the h-Mn ground state \cite{Besombes2014}. Values of $I_{hMn}$, $I_{eMn}$, $\rho_s/\Delta_{lh}$ and $\eta$ are obtained by comparing the linear polarization dependence of the experimental PL data to the optical transition probabilities calculated with the effective spin model (Fig.~\ref{Fig1}) \cite{Varghese2014}.

The Mn also exhibits a fine structure dominated by a magnetic anisotropy with an easy axis along the QD axis \cite{LeGall2009,LeGall2010}. Neglecting the small tetrahedral crystal field \cite{Causa1980,Qazzaz1995}, this fine structure is described by the spin Hamiltonian ${\cal H}_{Mn,CF}=D_0S^2_z+E(S_y^2-S_x^2)$. Here, $D_0$ is proportional to the biaxial strain and $E$ describes the anisotropy of the strain in the plane of the QD. $D_0$ varies from 0 $\mu$eV for a strain free QD \cite{Besombes2014} to 12 $\mu$eV for a fully strained CdTe layer matched on a ZnTe substrate \cite{LeGall2009}. These crystal field terms do not affect the PL spectra as they are present both in the excited (X$^{+}$-Mn) and ground (h-Mn) states. Nevertheless, we demonstrate here that ${\cal H}_{Mn,CF}$ drives the coherent dynamics of the e-Mn spin system.

The relaxation channels of the spin of X$^+$-Mn have been investigated by resonant PL and PL excitation (PLE). The PLE intensity map (inset of Fig.~\ref{Fig2}(a)) is obtained by scanning the high energy side of the X$^+$-Mn spectra while recording the PL from the low energy side. The three absorption resonances that are observed (labelled (1), (2) and (3)) give rise to the resonant PL spectra reported in Fig.~\ref{Fig2}(a). The PL is cross circularly polarized with the excitation except for an excitation on (1) which gives an unpolarized PL. The energy levels involved in the absorption resonances are identified and illustrated in Fig.~\ref{Fig2}(b). They correspond to the successive excitation of the e-Mn levels: $|3,+1\rangle$, $|3,+2\rangle$ and $|2,+2\rangle$. These states can be expressed as linear combinations of the Mn and electron spins $|S_z\rangle|\sigma_z\rangle$ coupled by a flip-flop :

\begin{tabular}{rl}
$|3,+1\rangle=$  &  $\frac{1}{\sqrt{6}}(\sqrt{4}|+1/2\rangle|\uparrow\rangle+\sqrt{2}|+3/2\rangle|\downarrow\rangle)$\\
$|3,+2\rangle=$ & $\frac{1}{\sqrt{6}}(\sqrt{5}|+3/2\rangle|\uparrow\rangle+\sqrt{1}|+5/2\rangle|\downarrow\rangle)$\\
$|2,+2\rangle=$ & $\frac{1}{\sqrt{6}}(\sqrt{1}|+3/2\rangle|\uparrow\rangle-\sqrt{5}|+5/2\rangle|\downarrow\rangle)$\\
\end{tabular}

Each of these e-Mn states is connected with circularly polarized optical transitions to two h-Mn ground states. For instance, $|3,+2\rangle$ is connected to $|+3/2\rangle|\Uparrow\rangle$  with $\sigma-$ light and to $|+5/2\rangle|\Downarrow\rangle$ with $\sigma+$ light. This level structure forms an optical $\Lambda$ system. Under resonant excitation of one high energy level of X$^+$-Mn, only one cross-circularly emission line is observed. It corresponds to the optically allowed recombination on the second branch of the $\Lambda$ system. This recombination occurs with a flip-flop of the electron and Mn spins \cite{Varghese2014}. The absence of PL on the low energy line under tunable excitation on the hight energy side of X$^+$-Mn (Fig.\ref{Fig2}(a)) is also consistent with this flip-flop process. This low energy transition corresponds to the e-Mn state $|3,+3\rangle=|+5/2\rangle|\uparrow\rangle$ in $\sigma-$ polarization and $|3,-3\rangle=|-5/2\rangle|\downarrow\rangle$ in $\sigma-$ polarization which have only one possible optical recombination or excitation channel. These resonant experiments show that the e-Mn and h-Mn states forms an ensemble of optical $\Lambda$ systems that can be addressed with a circularly polarized laser field.

\begin{figure}[hbt]
\includegraphics[width=3.25in]{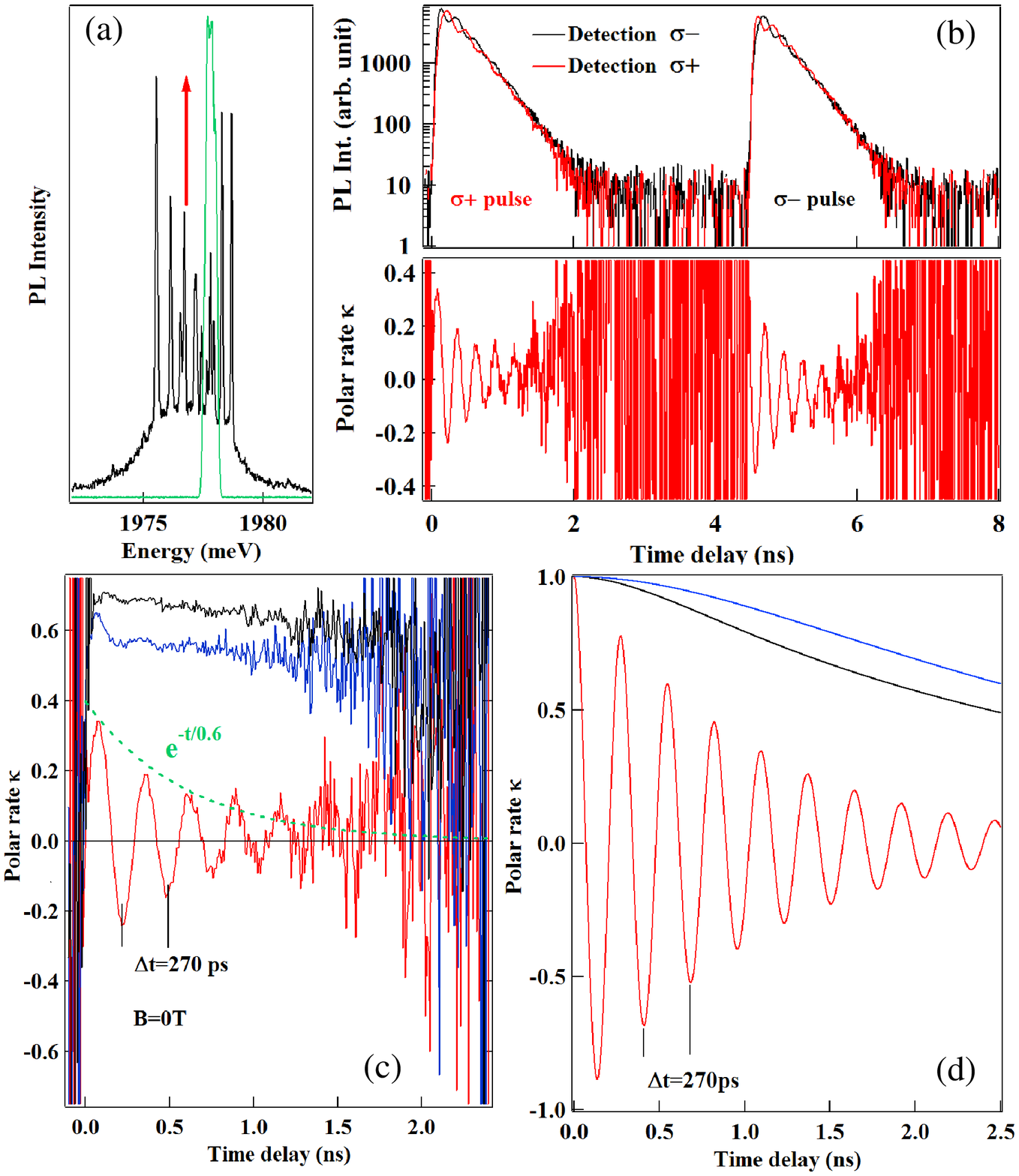}
\caption{(a) Configuration of the time resolved PL experiment for an excitation of $|3,+1\rangle$ (pulsed laser in green). (b) Top panel: Time resolved resonant PL of $|3,+1\rangle$ with a $\sigma+$/$\sigma-$ sequence of laser pulses and a detection in $\sigma+$ and $\sigma-$ polarization. Bottom panel: corresponding time dependence of the circular polarization rate $\kappa=(\sigma_{-}-\sigma_{+})/(\sigma_{-}+\sigma_{+})$. (c) Time dependence of the circular polarization rate of the resonant PL of the states $|3,+1\rangle$ (red), $|3,+2\rangle$ (black) and $|2,+2\rangle$ (blue). (d) Corresponding polarisation rates calculated with $D_0=7 \mu eV$ \cite{Varghese2014}, $T_2^{eMn}=0.6ns$, $E=1.8\mu eV$, a radiative lifetime $T_r=0.3ns$ and the parameters deduced from PL (Fig.\ref{Fig1}).}
\label{Fig3}
\end{figure}

We exploited this $\Lambda$ level structure to analyze the coherent dynamics of the e-Mn spin through the time evolution of the circular polarization rate, $\kappa=(\sigma_{Cross}-\sigma_{Co})/(\sigma_{Cross}+\sigma_{Co})$, of the resonant PL. The configuration of the experiment is summarized in figure \ref{Fig3}(a). Circularly polarized and spectrally filtered 10 ps laser pulses are successively tuned on resonance with the three absorption lines identified in the continuous wave experiment. This corresponds to independent optical excitation of the e-Mn states $|3,+1\rangle$, $|3,+2\rangle$ and $|2,+2\rangle$. The QD is excited with sequences of $\sigma+$/$\sigma-$ pulses (Fig.\ref{Fig3}(b)), to avoid any possible optical spin pumping of h-Mn\cite{Varghese2014} that could influence the observed dynamics.

The main result is the observation of an oscillatory behavior of the polarization rate of the PL when probing the dynamics of the $|3,+1\rangle$ state. The period of the beats is 270 ps with a characteristic damping time of 0.6 ns. When probing the dynamics of the $|3,+2\rangle$ and $|2,+2\rangle$ states, we measured cross circularly polarized PL with a slow decrease of the polarization rate during the lifetime of X$^+$-Mn.

The origin of this dynamics lies in the fine structure of the e-Mn levels. The $|3,+1\rangle$ and $|3,-1\rangle$ states are degenerated and differ by a change of angular momentum of two. Consequently, they are efficiently mixed by the anisotropic strain term $E(S_y^2-S_x^2)$ which induces a spin-flip of two of the Mn with a conservation of the electron spin. This coupling has no significant influence on the other e-Mn states which are initially split by $D_0 S_z^2$ and $-2\eta S_z^2$.

When a pulsed laser is tuned to the high energy transition of the $\Lambda$ system associated to $|3,+1\rangle$ ($\sigma+$ absorption from the h-Mn state $|+3/2\rangle|\Downarrow\rangle$), the PL of the low energy transition of the $\Lambda$ system is first cross-circularly polarized ($\sigma-$ recombination to the h-Mn state $|+1/2\rangle|\Uparrow\rangle$). Then, after a coherent transfer of population to the e-Mn state $|3,-1\rangle=\frac{1}{\sqrt{6}}(\sqrt{2}|-3/2\rangle|\uparrow\rangle+\sqrt{4}|-1/2\rangle|\downarrow\rangle)$, induced by $E$ (Fig.\ref{Fig1}(c)), co-circularly polarized PL is emitted at the same energy from the $|3,-1\rangle$ state ($\sigma+$ recombination to $|-1/2\rangle|\Downarrow\rangle$). This coherent transfer of population is fully controlled by the in-plane anisotropy of the strain at the Mn location and is responsible for the observed oscillations of the circular polarization rate.

To understand the details of this dynamics, we calculated the time evolution of the populations and coherence of the twelve X$^+$-Mn states and the twelve hole-Mn states. We neglected here the hyperfine coupling between the electronic and nuclear spins of the Mn and solved the master equation for the 24 x 24 density matrix numerically, including relaxation and pure dephasing processes in the Lindblad form \cite{Exter2009,Roy2011,Jamet2013}. At zero magnetic field, the coherent evolution of this multi-level system is controlled by ${\cal H}_{Mn,CF}+{\cal H}_{eMn}^{ex}+2{\cal H}_{scat}$ for $X^+-Mn$ and ${\cal H}_{Mn,CF}+{\cal H}_{hMn}^{ex}+{\cal H}_{scat}$ for h-Mn \cite{supp}.

\begin{figure}[hbt]
\includegraphics[width=3.4in]{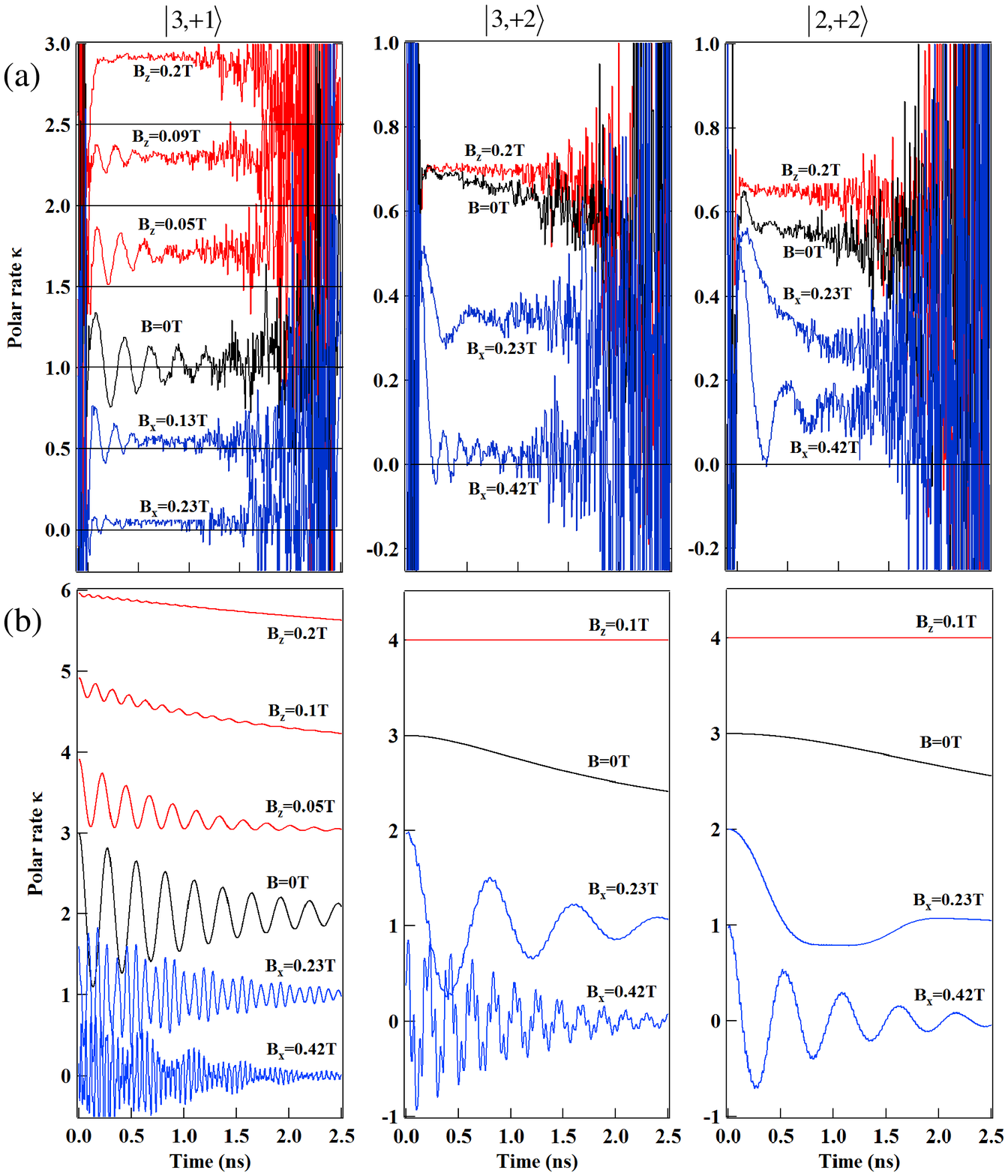}
\caption{(a) Influence of a longitudinal (B$_z$, red) and a transverse (B$_x$, blue) magnetic field on the time dependence of the circular polarization rate $\kappa=(\sigma_{-}-\sigma_{+})/(\sigma_{-}+\sigma_{+})$ of the resonant PL of $|3,+1\rangle$, $|3,+2\rangle$ and $|2,+2\rangle$. On the top left panel, curves are shifted by 0.5 for clarity. (b) Corresponding time dependence of the circular polarization rate calculated with $g_{Mn}=2$, $g_{e}=-0.4$, $g_{h}=0.6$ \cite{Varghese2014}, and the parameters of Fig.~(\ref{Fig3}). The curves are shifted by 1 for clarity.}
\label{Fig4}
\end{figure}

In the absence of magnetic field, the period of the quantum beats observed for an excitation of $|3,+1\rangle$ depends only on the anisotropy term $E$. The experimental data can be well reproduced by the model with $E=1.8\mu eV$ (Fig.\ref{Fig3}(d)). A coherence time, $T_2^{eMn}\approx0.6ns$, of the spin of e-Mn is extracted from the damping of the oscillations. For an excitation of $|3,+2\rangle$ and $|2,+2\rangle$ one can observe a slow decrease of the polarization rate which is also qualitatively reproduced by the model.

The coherent transfer of population depends both on the initial splitting of the e-Mn spin states (controlled at zero field by $D_0$ and $\eta$) and on the strength of the coupling $E$. The splitting between the e-Mn states can be tuned by a magnetic field, $B_z$, applied along the growth axis. In addition, a coupling between the e-Mn spin states $M_z$ can be induced by a magnetic field, $B_x$, applied in the QD plane. The experimental and calculated evolution of the polarization rate of the e-Mn states, $|3,+1\rangle$, $|3,+2\rangle$ and $|2,+2\rangle$, versus magnetic field are presented in Fig.~\ref{Fig4}.

Under a longitudinal magnetic field B$_z$, the e-Mn states M$_z=\pm1$ are split and the influence of $E$ is progressively reduced. For an excitation on the $|3,+1\rangle$ state, the amplitude and period of the oscillations in the polarization rate reduce as $B_z$ increases: The resonant PL becomes cross-circularly polarized with a polarization rate constant during the lifetime of $X^+$. A weak longitudinal magnetic field stabilizes the spin of the e-Mn states $|3,+2\rangle$ and $|2,+2\rangle$ and their polarization rate remains constant during the lifetime of X$^+$-Mn.

In a transverse magnetic field B$_x$, the quantum beats observed for an excitation of $|3,+1\rangle$ are accelerated and the measured circular polarization rate drops to zero as the period of the oscillations becomes smaller than the time resolution of the experimental setup ($\approx$60 ps). A given transverse magnetic field induces a slower oscillation of the polarization rate for the states $|3,+2\rangle$ and $|2,+2\rangle$.

The observed magnetic field dependence of the coherent dynamics of $|3,+1\rangle$, $|3,+2\rangle$ and $|2,+2\rangle$ can be qualitatively reproduced by the model with the exchange parameters deduced from the PL and the strain anisotropy term and coherence time deduced from the oscillations observed on $|3,+1\rangle$ at zero magnetic field (Fig.\ref{Fig3}). D$_0$ cannot be extracted from these measurements and we use a typical value D$_0$=7 $\mu eV$ corresponding to a partial relaxation of the biaxial strain \cite{Varghese2014}. The different precession periods observed for the three states in a given transverse magnetic field are particularly well described \cite{supp}.

To conclude, we have demonstrated that, in p-doped magnetic QDs, the hole-Mn states and the positively charged exciton levels ({\it i.e.} e-Mn levels) form an ensemble of optical $\Lambda$ systems. This opens the possibility to perform a coherent manipulation of the magnet formed by the coupled h-Mn spins with two resonant optical fields \cite{Houel2014}. We have shown that the strain amplitude and symmetry at the Mn location controls the coherent dynamics of the coupled electron and Mn spins. These results demonstrate the potential of magnetic QDs where one could exploit the intrinsic spin-strain interaction to coherently couple the spin of a magnetic atom to the motion of a nano-mechanical oscillator \cite{Kolkowitz2012,Teissier2014} and suggest some possible coherent mechanical spin-driving of a magnetic atom.

\section*{Supplementary information to "Strain induced coherent dynamics of coupled carriers and Mn spins in a quantum dot"}

In this supplementary material, we present in details the model used for the calculation of the X$^+$-Mn coherent dynamics under longitudinal or transverse magnetic field. We show in particular the influence of the in-plane orientation of the transverse magnetic field on the electron-Mn dynamics.

\subsection*{X$^+$-Mn energy levels}

\begin{figure}[hbt]
\includegraphics[width=3.35in]{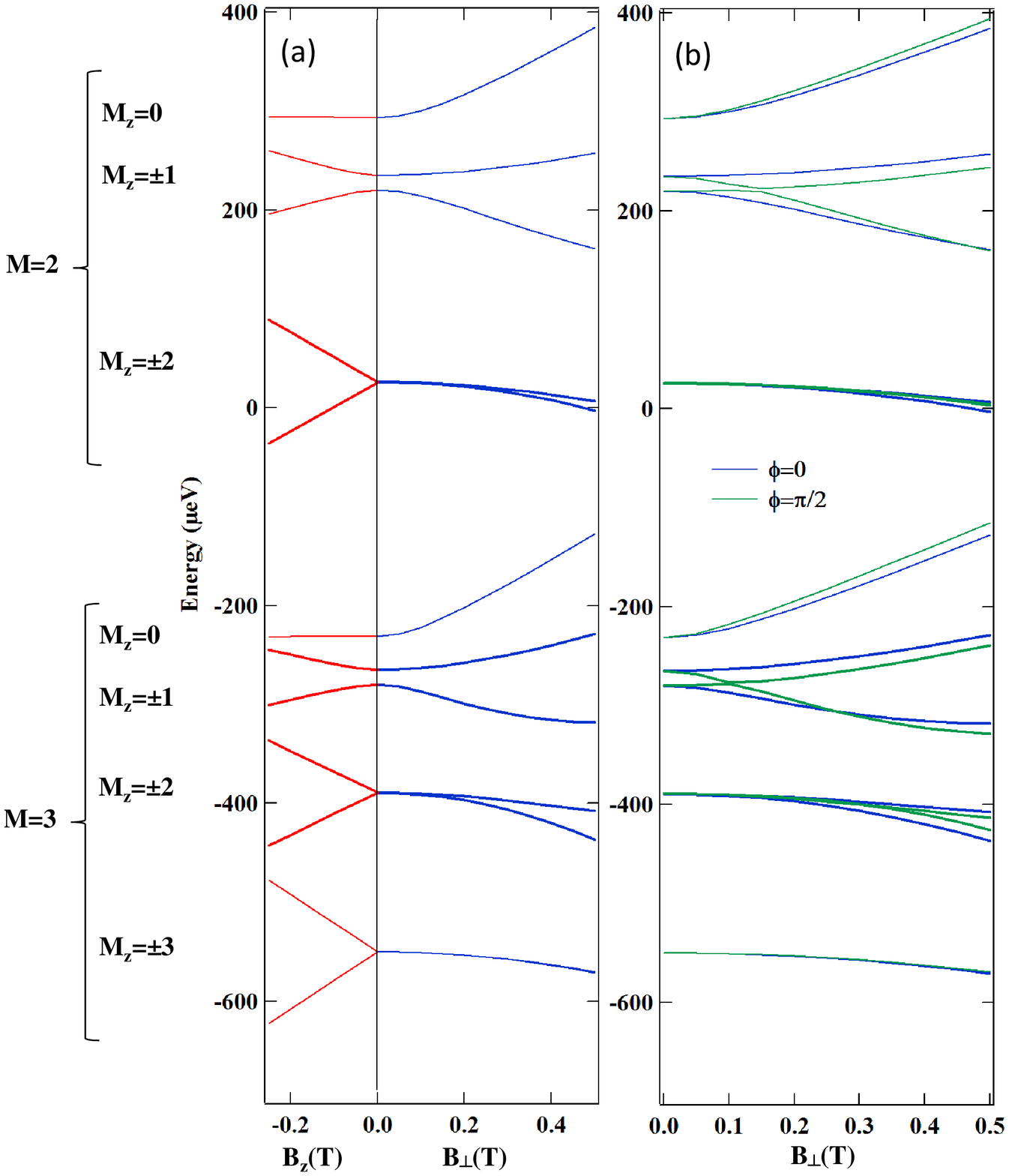}
\caption{(Color online) (a) Calculated energy of the electron-Mn states in a longitudinal magnetic field (B$_z$) and in a transverse magnetic field (B$_{\perp}$=B$_x$). (b) Energy of the electron-Mn states for two orientations of the transverse magnetic field $\phi$=0 (B$_{\perp}$=B$_x$) and $\phi=\pi/2$ (B$_{\perp}$=B$_y$). The parameters used in the calculation are $I_{eMn}=-175\mu eV$, $\eta=30\mu eV$, $D_0=7 \mu eV$, $E=1.8\mu eV$, g$_{Mn}$=2 and g$_e$=-0.4.}
\label{FigS1}
\end{figure}

The coherent dynamics of X$^+$-Mn is mainly controlled by the electron-Mn Hamiltonian. The magnetic field dependence of the electron-Mn energy levels calculated with the Hamiltonian

\begin{eqnarray}
{\cal H}_{e-Mn}=I_{eMn}\vec{S}\cdot\vec{\sigma}-2\eta S_z^2+D_0S^2_z+E(S_y^2-S_x^2)\nonumber\\
+g_{Mn}\mu_B\vec{S}\cdot\vec{B}+g_{e}\mu_B\vec{\sigma}\cdot\vec{B}
\end{eqnarray}

\noindent are presented in Fig.~\ref{FigS1} for a longitudinal (B$_z$) and transverse (B$_\bot$) magnetic field.

At zero magnetic field, the electron-Mn states M=2 on one hand and M=3 on the other hand are split by the strain induced magnetic anisotropy $D_0 S_z^2$ and the interaction with the two holes $-2\eta S_z^2$. In addition, the strain anisotropy term $E(S_y^2-S_x^2)$ couple the states M$_z$=$\pm$1 and the new eigenstates are split by an energy which only depend on the strain distribution at the Mn atom location. In a longitudinal magnetic field, the electron-Mn states M$_z$=$\pm$i are split by the Mn Zeeman energy.

In a transverse magnetic field, the observed splitting of the states M$_z$=$\pm$i strongly depend on the considered levels. The field induced splitting results in this case from mixing of electron-Mn states with a difference in total angular momentum $\Delta M_z=\pm1$. These states are initially irregularly separated at zero field resulting in different transverse magnetic field induced coupling. The coherent dynamics in transverse magnetic field presented in figure 4 of the paper is a consequence of this magnetic field dependence of the energy levels.

As presented in Fig.~\ref{FigS1}(b), the energy of the electron-Mn levels also depend on the angle of the transverse magnetic field in the plane of the quantum dot (QD). The splitting induced by the anisotropy of the strain described by $E(S_y^2-S_x^2)$ has a preferential axis and can be cancelled for a given in-plane direction and amplitude of the transverse magnetic field.

\subsection*{X$^+$-Mn spin dynamics}

To model the polarization rate of the photoluminescence of X$^+$-Mn under pulsed resonant excitation, we calculate the time evolution of the population and coherence of the twelve X$^+$-Mn states in the excited state of the QD and twelve hole-Mn states in the ground state. We solve the master equation for the 24 x 24 density matrix $\rho$ numerically, including relaxation and pure dephasing processes in the Lindblad form:

\begin{equation}
\frac{\partial \rho}{\partial t}=-i/\hbar[{\cal H},\rho]+L\rho
\end{equation}

\noindent where ${\cal H}$ is the Hamiltonian of the complete system ($X^+$-Mn and h-Mn) and $L\rho$ describes the coupling or decay channels resulting from an interaction with the environment \cite{Exter2009,Roy2011,Jamet2013}. The coherent evolution of this multi-level system is controlled by the Hamiltonian ${\cal H}_{e-Mn}$ for X$^+$-Mn in the excited state of the QD and

\begin{eqnarray}
{\cal H}_{h-Mn}=I_{hMn}\vec{S}\cdot\vec{J}-\eta S_z^2+D_0S^2_z+E(S_y^2-S_x^2)\nonumber\\
+g_{Mn}\mu_B\vec{S}\cdot\vec{B}+g_{h}\mu_B\vec{J}\cdot\vec{B}
\end{eqnarray}

\noindent for the h-Mn complex in the ground state.

\begin{figure}[hbt]
\includegraphics[width=3.35in]{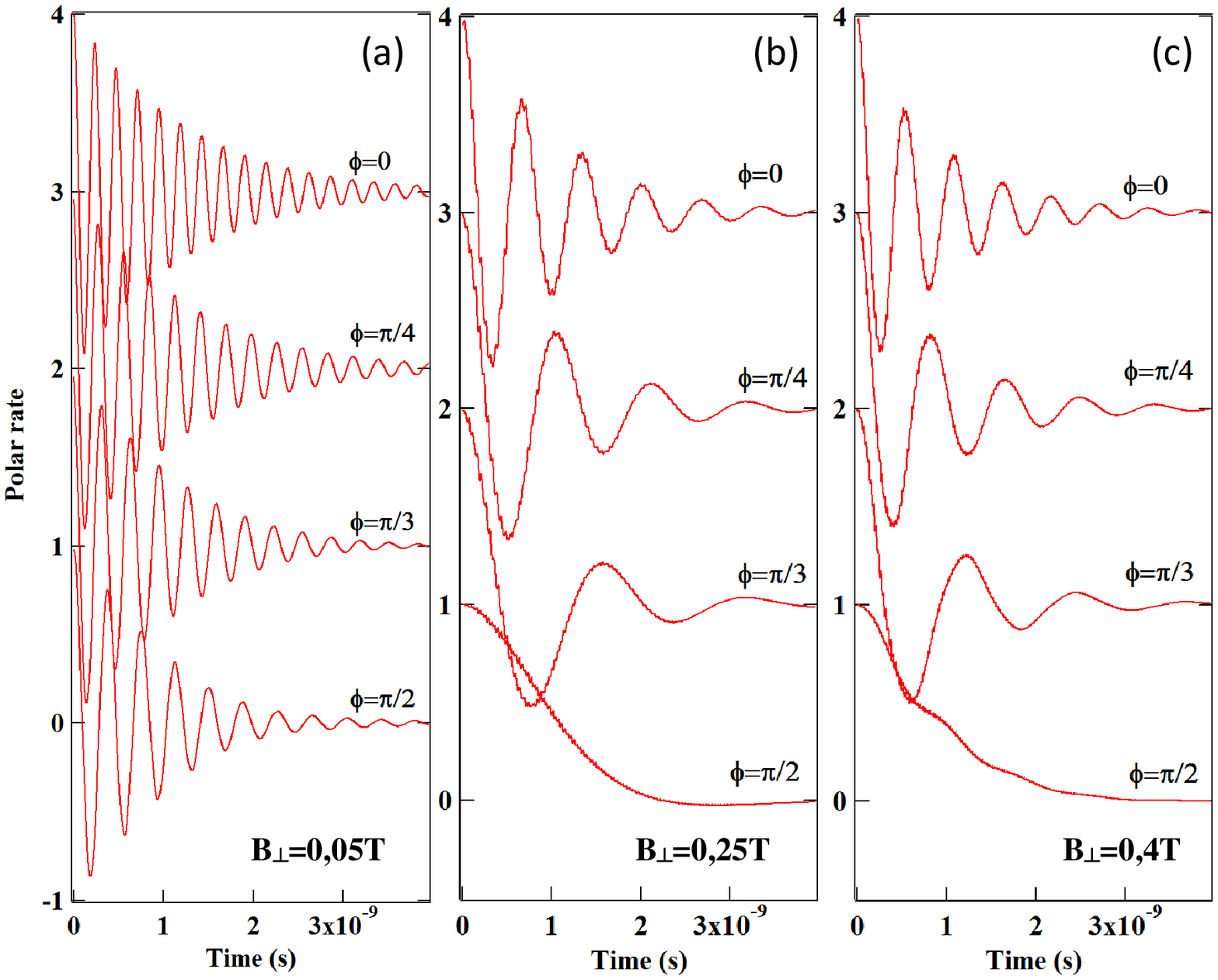}
\caption{(Color online) Calculated time dependence of ($\rho_{|+1/2\rangle|\uparrow\rangle}$-$\rho_{|-1/2\rangle|\downarrow\rangle}$)/($\rho_{|+1/2\rangle|\uparrow\rangle}$+$\rho_{|-1/2\rangle|\downarrow\rangle}$) for a pulsed excitation from $|+3/2\rangle|\Downarrow\rangle$ to the subspace M=3 (a), calculated time dependence of ($\rho_{|+3/2\rangle|\uparrow\rangle}$-$\rho_{|-3/2\rangle|\downarrow\rangle}$)/($\rho_{|+3/2\rangle|\uparrow\rangle}$+$\rho_{|-3/2\rangle|\downarrow\rangle}$) for an excitation from $|+5/2\rangle|\Downarrow\rangle$ to the subspace M=3 (b) and time dependence of ($\rho_{|+3/2\rangle|\uparrow\rangle}$-$\rho_{|-3/2\rangle|\downarrow\rangle}$)/($\rho_{|+3/2\rangle|\uparrow\rangle}$+$\rho_{|-3/2\rangle|\downarrow\rangle}$) for an excitation from $|+5/2\rangle|\Downarrow\rangle$ to the subspace M=2 (c) for different in-plane directions of a fixed transverse magnetic field. The parameters used in the calculation are $I_{hMn}=345 \mu eV$, $I_{eMn}=-175\mu eV$, $\rho_s/\Delta_{lh}=0.09$, $\eta=30\mu eV$, $D_0=7 \mu eV$, $E=1.8\mu eV$, $T_2^{eMn}=0.6ns$, $T_r=0.3ns$, $g_{Mn}=2$, $g_{e}=-0.4$ and $g_{h}=0.6$. The spin relaxation on the hole-Mn system (ground state of the QD) in the $\mu s$ range is longer than the timescale considered here and we use $T^{hMn}=\infty$.}
\label{FigS2}
\end{figure}

For the initial condition in the calculation of the time evolution, we consider that a $\sigma+$ pulse on resonance with the absorption line (1) (see figure 2 of the paper for identification of the absorption lines) projects the system on the M=3 electron-Mn subspace on all the levels that contain a component $|+3/2\rangle|\downarrow\rangle$. In the absence of transverse magnetic field and strain anisotropy term $E$, this excitation simply corresponds to an optical transition from the hole-Mn state $|+3/2\rangle|\Downarrow\rangle$ towards the electron-Mn state $|3,+1\rangle$. With a weak transverse magnetic field (typically lower than 0.5 T), a linear combination of the M=3 states is created. At large transverse magnetic field, one should consider possible mixing with the M=2 states. Similarly, a $\sigma+$ pulse on (2) projects the system on the M=3 electron-Mn subspace on the levels that contain a component $|+5/2\rangle|\downarrow\rangle$ and a $\sigma+$ pulse resonant on (3) projects the system on the M=2 electron-Mn subspace on the levels that contain a component $|+5/2\rangle|\downarrow\rangle$.

After this excitation, the circular polarization of the resonant photoluminescence is governed by the evolution of the spin of the electron. For instance, to compute the circular polarization rate of the emission after a resonant $\sigma+$ excitation on (1) (optical excitation from the hole-Mn state $|+3/2\rangle|\Downarrow\rangle$ to $|3,+1\rangle$: high energy branch of the $\Lambda$-system) we calculate the difference between the density matrix elements $\rho_{|+1/2\rangle|\uparrow\rangle}$ ($\sigma-$ recombination towards the hole-Mn state $|+1/2\rangle|\Uparrow\rangle$: low energy branch of the $\Lambda$-system) and $\rho_{|-1/2\rangle|\downarrow\rangle}$ ($\sigma+$ recombination towards the hole-Mn state $|-1/2\rangle|\Downarrow\rangle$: low energy branch of the $\Lambda$-system associated with $|3,-1\rangle$).

The calculated time dependence of the circular polarization rate in transverse magnetic field are presented in Fig.~\ref{FigS2} for the three $\Lambda$-systems identified in figure 2 of the paper. In agreement with the calculated electron-Mn energy levels (Fig.\ref{FigS1}), this modelling reveals a significant influence of the orientation of the transverse magnetic field on the electron-Mn coherent dynamics. This behaviour could not be observed with our experimental set-up which did not permit to change the direction of the transverse field while staying on the same Mn-doped QD. A systematic experimental study of the transverse magnetic field effect has still to be realized.

\end{document}